\documentclass[prl,twocolumn,groupedaddress]{revtex4}
\usepackage{graphicx}
\usepackage{amsmath}
\usepackage{amssymb}
\usepackage{bm}
\usepackage{color}
\usepackage{hyperref}
\usepackage{tabularx}
\usepackage{multirow}
\usepackage{braket}
\usepackage{epstopdf}

\begin{document}
\title{Supercurrent diode effect 
and finite momentum superconductivity
}
	
\author{{Noah F. Q. Yuan$^{1,2}$}}
\thanks{These two authors contribute equally to this work}
\email{fyuanaa@connect.ust.hk}
\author{{Liang Fu$^{2*}$}}
\email{liangfu@mit.edu}
\affiliation{1. Shenzhen JL Computational Science and Applied Research Institute, Shenzhen, 518109 China\\
2. Department of Physics, Massachusetts Institute of Technology, Cambridge,
Massachusetts 02139, USA}

\begin{abstract}
When both inversion and time-reversal symmetries are broken, the critical current of a superconductor can be nonreciprocal. In this work we show that in certain classes of two-dimensional superconductors with antisymmetric spin-orbit coupling, Cooper pairs acquire a finite momentum upon the application of an in-plane magnetic field, and as a result, critical currents in the direction parallel and antiparallel to the Cooper pair momentum become unequal. This supercurrent diode effect is also manifested in the polarity-dependence of in-plane critical fields induced by a supercurrent.  
These nonreciprocal effects may be found in polar SrTiO$_3$ film, few-layer MoTe$_2$ in the $T_d$ phase, and twisted bilayer graphene in which the valley degree of freedom plays the role analogous to spin.
\end{abstract}

\maketitle

\textbf{Introduction}--- Shortly after the advent of Bardeen-Cooper-Schrieffer (BCS) theory of superconductivity, it was predicted that a superconducting phase with a spatially varying order parameter exists in a narrow range of magnetic fields above the Pauli limit \cite{FF,LO}. In this Fulde-Ferrell-Larkin-Ovchinnikov (FFLO) phase, pairing of opposite spin states on Zeeman-split Fermi surfaces gives rise to finite Cooper pair momentum. 
Extensive efforts have been devoted to search for FFLO phase in clean superconductors. 
Thermodynamic and NMR measurements have found evidence of a distinctive superconducting phase at high fields in several materials \cite{CeCoIn5-1,CeCoIn5-2,CeCoIn5-3,CeCu2S2,Organic1,Organic2,Organic3,KFe2As2,FeSe,Joe}. However, the existence of finite Cooper pair momentum has not been demonstrated directly.

Recently, a new type of finite-momentum superconducting state has been predicted in two-dimensional systems with strong spin-orbit coupling (SOC) and broken inversion symmetry \cite{Gorkov,MichaeliPotterLee,Samokhin1,Samokhin2,Edelstein1,Edelstein2,Agterberg,Kaur,YuanPNAS,Aoyama,MWWu,Olga1,Olga2,Buzdin,Fulde}. Here SOC splits Fermi surfaces and creates (topologically) nontrivial spin textures in momentum space \cite{YuanPNAS, Fu}.     
Upon the application of a parallel magnetic field, a BCS superconductor with spin-textured Fermi surfaces 
can {\it smoothly} evolve into a finite momentum state with 
the phase of the superconducting order parameter being periodically modulated as
$\Delta({\bm r})=\Delta e^{i {\bm q} \cdot {\bm r}}$, where $\bm q$ is induced by and varies {\it continuously} with the magnetic field.
This state is formed by pairing {\it within} each spin-non-degenerate Fermi surface. {Such helical superconducting phase should be distinguished from the single-$\bm q$ FFLO (helical FF) state that 
is separated from the BCS state by a first-order transition.
Possible realization of helical superconductivity has been proposed in several noncentrosymmetric materials \cite{MichaeliPotterLee, Kaur, Matsuda}}. However, 
it has been unclear how to detect this state easily and unambiguously. 

\begin{figure}[ht]
\begin{center}
\leavevmode\includegraphics[width=1\hsize]{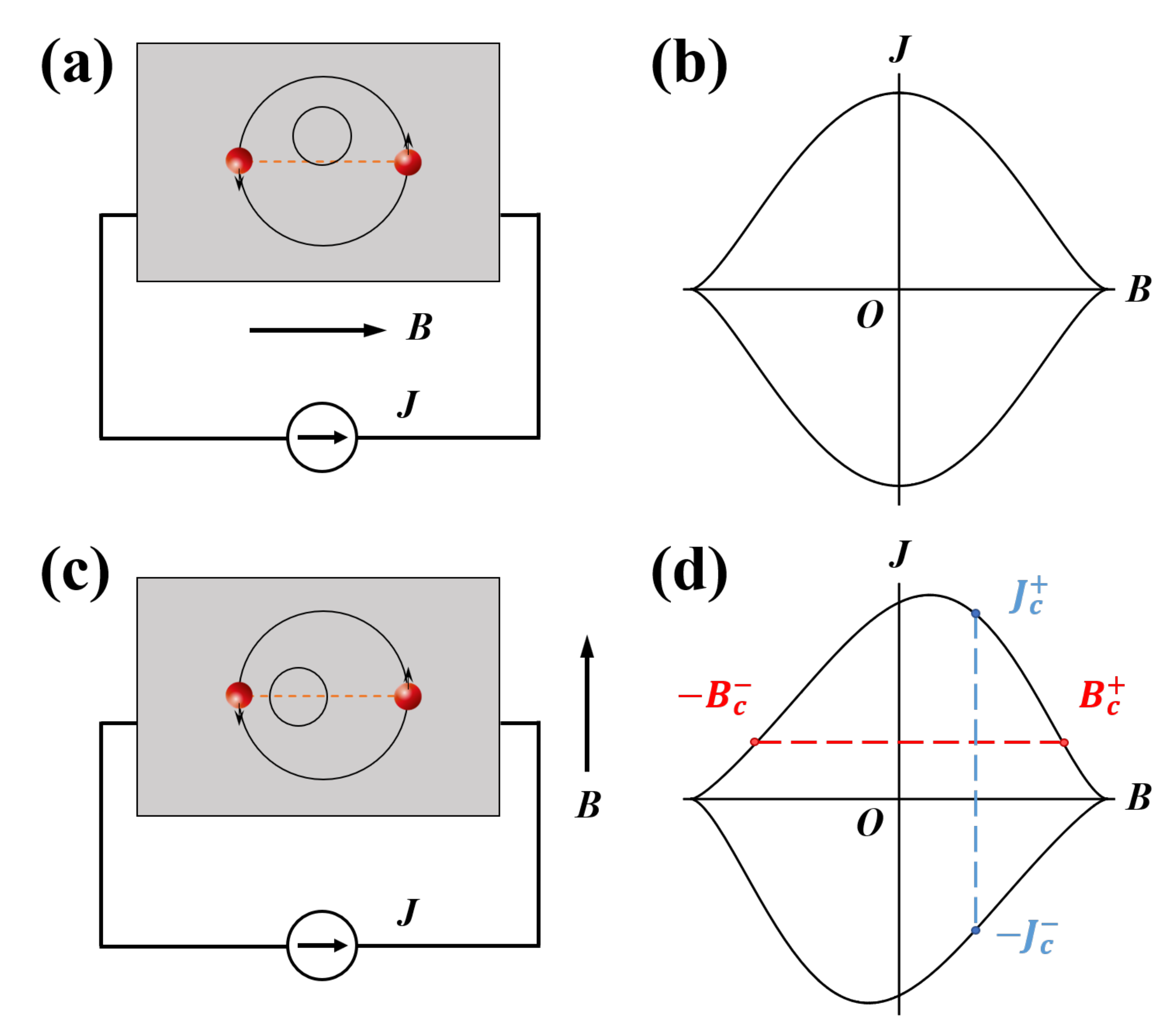}
\end{center}
\caption{Supercurrent diode effect in a Rashba superconductor under in-plane magnetic field $B$ and external current source $J$. Panels (a,c) are device plots with circles denoting normal state Fermi surfaces, and (b,d) denote schematic phase diagram in $B$-$J$ plane. When $\bm B\parallel\bm J$ in (a), the phase diagram in (b) is symmetric with respect to both $B$ and $J$ axes. And when $\bm B\perp\bm J$ in (c), the phase diagram in (d) is skewed, indicating nonreciprocal critical current $J_{c}^{+}\neq J_{c}^{-}$ and polarity-dependent critical field $B_{c}^{+}\neq B_{c}^{-}$.}
\label{fig_0}
\end{figure}

In this work, we show that helical superconductors exhibit an intrinsic supercurrent diode effect: 
the depairing critical currents in the direction along and against the underlying Cooper pair momentum ${\bm q}_0$ are different, as shown in Fig. \ref{fig_0}. This effect is also manifested in the polarity-dependent in-plane critical fields in the presence of a supercurrent. These nonreciprocal phenomena are a direct consequence of the Cooper pair momentum which breaks time-reversal and inversion symmetry  in the equilibrium state.

Our work is motivated by the recent observation of nonreciprocal critical current in an 
artificial metal film under a parallel magnetic field \cite{TeruoNature}. Its origin is not yet fully understood 
and likely due to the orbital effect associated with the film thickness \cite{FuCommentary}. We also note recent works showing that superconducting fluctuations  enhance nonreciprocal resistance above $T_c$ \cite{Nagaosa1, Nagaosa2} as well as works on nonreciprocal Josephson current across a weak link \cite{Vic,Molenkamp,Ong,Buzdin2,Reynoso,Zazunov,Nagaosa,Mohammad,Silaev}. 
Unlike these previous works, our focus is thin two-dimensional superconductors where the orbital effect is suppressed, and our study provides for the first time 
a microscopic theory of intrinsic supercurrent diode effect 
in superconductors.

\textbf{Helical superconductivity}--- We consider a 2D electron system with spin-orbit coupling (SOC) under in-plane magnetic field
\begin{equation}\label{eq_d1}
    H_{\bm k}=\xi_{\bm k} +\bm g_{\bm k}\cdot\bm\sigma +\bm B\cdot\bm\sigma,
\end{equation}
where $ \bm k=(k_x,k_y) $ is the 2D momentum, $\xi_{\bm k}=\frac{k^2}{2m}-\mu$ is the kinetic term, $m$ is effective mass, $\mu$ is chemical potential, Pauli matrices $ \bm\sigma =(\sigma_x,\sigma_y,\sigma_z) $ denote spin, $\bm g_{\bm k}$ is the SOC vector and $\bm B$ is the Zeeman energy due to the in-plane magnetic field.
Such 2D electron gas has two spin-split energy bands 
\begin{eqnarray}
\xi^{\pm}_{\bm k}= \xi_{\bm k}\pm |\bm g_{\bm k}+\bm B|
\end{eqnarray} 
and hence two spin-non-degenerate Fermi surfaces.

The location, shape and spin configuration of both Fermi surfaces evolve with the in-plane magnetic field. We will take Rashba SOC $\bm g_{\bm k}=\alpha_{\rm R}\hat{\bm z}\times\bm k$ as an example \cite{YuanPNAS}.
At zero field, two concentric Fermi circles are centered at ${\bm k}=0$, with helical spin textures. These two Fermi surfaces have different density of states (DOS) $ N_{\pm}=\frac{1}{2}N_0(1\mp\alpha_{\rm R}/\overline{v}) $ ($\overline{v}=\sqrt{v^2_{\rm F}+\alpha_{\rm R}^2}$), where $N_0=4\pi m$ is the total DOS including spin degeneracy and $v_{\rm F}=\sqrt{2\mu/m}$ is the Fermi velocity. A small field $B\ll\Delta_{so}\equiv m\alpha_{\rm R}v_{\rm F}$ displaces the centers of inner $(+)$ and outer $(-)$ Fermi pockets to opposite momenta $\pm{\bm k}_0=\pm\hat{\bm z}\times\bm B/v_{\rm F}$ respectively, as shown in Fig. \ref{fig_0}a and c. To the first order in $B$, 
the energy dispersion satisfies  
$
\xi^+_{\bm k + {\bm k}_0} = \xi^+_{-\bm k + {\bm k}_0}, \;
\xi^-_{\bm k - {\bm k}_0} = \xi^-_{-\bm k - {\bm k}_0}. 
\label{BCScondition}
$
Therefore each  Fermi surface remains nearly symmetric with respect to its displaced center, and the spin configuration remains nearly helical.  
 


At $B=0$, short-range attractive interaction leads to a BCS superconductor with zero momentum pairing, where states at $\pm {\bm k}$ of opposite spins within each Fermi pocket are paired.   
At small $B$, the approximate inversion symmetry of the outer (inner) Fermi pocket with respect to its displaced center $\mp {\bm k}_0$ naturally favors BCS-type intra-pocket pairing, which leads to a nonzero Cooper pair momentum $\mp 2 {\bm k}_0$. 
Since the outer pocket has larger DOS, the state with Cooper pair momentum $\bm q_0 \approx -2\bm k_0$ is energetically favored. It has an isotropic gap on the outer Fermi pocket {(see Eq.(\ref{eq_e}) below)}, whereas the gap on the inner pocket is anisotropic due to the combined pair breaking effect of Zeeman field and Cooper pair momentum. {As discussed in the Supplemental Information, the competition between $\mp 2 {\bm k}_0$ Cooper pairs can also lead to other phases, and in the rest of this manuscript we will focus on the single-$\bm q$ helical state unless specified otherwise.} 

Such helical state induced by Zeeman and SOC effects at small $B$  is smoothly connected to the ${\bm q}=0$ BCS state in the limit $B=0$. Provided that  
the SOC strength $\Delta_{so}$ is much larger than the BCS pairing gap $\Delta_0$ at $B=0$,  
the field-induced helical state persists in the strong disorder regime 
$\Delta_0 \ll \tau^{-1} \ll \Delta_{so}$ \cite{MichaeliPotterLee}.
These properties of the helical superconductor clearly contrast  with the helical FF state that is formed by pairing between inner and outer pockets, is separated from BCS state by a first-order transition at the Pauli limiting field, and is highly sensitive to disorder.  The plethora of 2D superconductors recently found in spin-orbit-coupled systems \cite{Iwasa2017review} provides unprecedented opportunity to find helical superconductivity.
However, it is difficult to distinguish a helical superconductor having a spatially uniform full gap from the BCS state. A direct measurement of the Cooper pair momentum requires sophisticated interference experiments using a Josephson junction between a helical superconductor and 
a reference BCS superconductor \cite{Kaur}.

\textbf{Origin of supercurrent diode effect}--- In this work we predict that as a direct consequence of nonzero Cooper pair momentum $\bm q_0$ 
in equilibrium state, helical superconductors generally exhibit nonreciprocal critical current: the maximum current that can flow with zero resistance in the direction along $\bm q_0$ differs from the one in the  opposite direction.    
Here we consider the depairing critical current, which is associated with 
the reduction and eventual closing of the superconducting gap with increasing supercurrent. 

\begin{figure}[ht]
\begin{center}
\leavevmode\includegraphics[width=1\hsize]{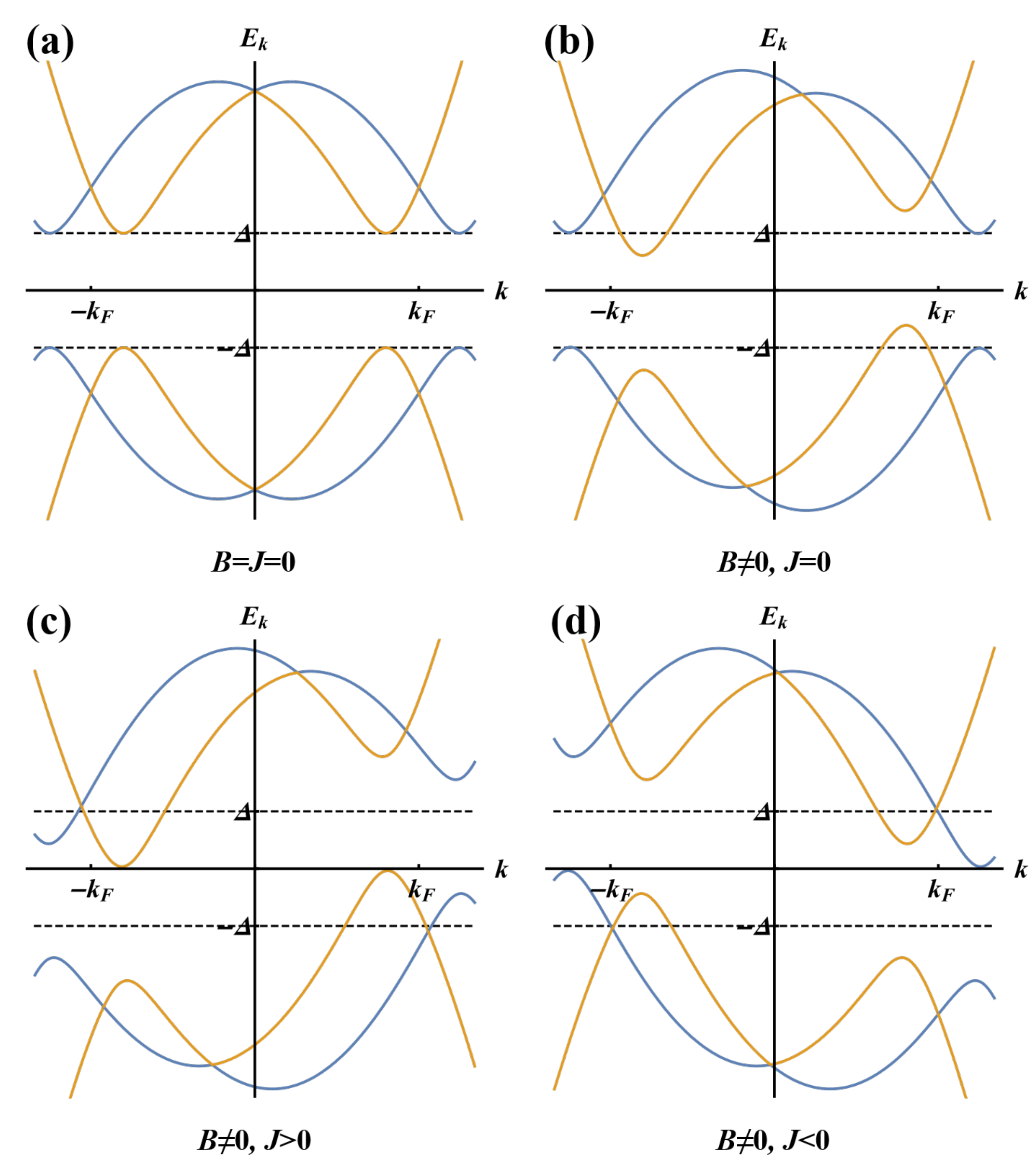}
\end{center}
\caption{Energy spectra of {superconductors with spin-orbit coupling:} (a) {conventional superconductor without external field and external current}, (b) helical superconductor {with external field} without external current and (c,d) {helical superconductor with external field and} external current along opposite directions. We use Eqs. (\ref{eq_d1}) {and (\ref{eq_d2})} with $m=1,\mu=10,\alpha_{\rm R}=1,\bm B=B\hat{\bm y},\bm J=J\hat{\bm x}$ and pairing potential $\Delta=3$. In (b-d) $B=0.6\ll\Delta_{so}=2\sqrt{5}$.}
\label{fig_1}
\end{figure}

The origin of nonreciprocal critical current in a helical superconductor 
can be understood heuristically from the gap structure. Assuming a local attractive 
interaction 
the 
mean-field Hamiltonian of the helical superconductor reads
\begin{equation}\label{eq_mf}
H_{\rm MF} = \sum_{\bm k} c_{\bm k}^\dagger H_{\bm k} c_{\bm k} +\sum_{\bm r} \Delta(\bm r) (c_{\bm r\uparrow}^\dagger c_{\bm r \downarrow} + h.c.)
=\frac{1}{2}\Psi_{\bm k}^{\dagger}\mathcal{H}_{\bm k}\Psi_{\bm k} 
\end{equation}
where $\Delta(\bm r)=\Delta e^{i {\bm q} \cdot {\bm r}}$ is the superconducting order parameter, $\Delta$ is the pairing potential, $\bm q$ is Cooper pair momentum,
$\Psi_{\bm k}=\left(c_{\bm k+\frac{1}{2}\bm q \uparrow}, c_{\bm k+\frac{1}{2}\bm q \downarrow}, c^\dagger_{-\bm k+\frac{1}{2}\bm q \uparrow}, c^\dagger_{-\bm k+\frac{1}{2}\bm q \downarrow} \right)^{\rm T}$ is the Nambu basis,
and the Bogouliubov-de Gennes Hamiltonian $\mathcal{H}_{\bm k}$ takes the form
\begin{equation}\label{eq_d2}
    \mathcal{H}_{\bm k}(\bm q,\Delta)=
    \begin{pmatrix}
    H_{\bm k+\frac{1}{2}\bm q} & -i\sigma_y \Delta \\
    i\sigma_y \Delta & -H^{*}_{-\bm k+\frac{1}{2}\bm q}
    \end{pmatrix}.
\end{equation}
By construction, the BdG Hamiltonian satisfies the anti-unitary particle-hole symmetry $\mathcal{P}\mathcal{H}^{*}_{\bm k}\mathcal{P}^{-1}=-\mathcal{H}_{-\bm k}$ where $\mathcal{P}=\tau_x$ acts in Nambu space that double counts the degrees of freedom. 
The spectrum of $\mathcal{H}_{\bm k}$ consists of pairs of opposite eigenvalues $E_{\lambda}({\bm k}), -E_{\lambda}({-\bm k})$, see Fig. \ref{fig_1}a-d, where $\lambda=\pm$ denotes quasiparticle states associated with the two spin-split energy bands of normal phase.
At $B=q=0$,  the presence of time-reversal symmetry leads to $\mathcal{T}\mathcal{H}^{*}_{\bm k}\mathcal{T}^{-1}=\mathcal{H}_{-\bm k}$ where $\mathcal{T}=i\sigma_y$ acts on spin, and the spectrum of $\mathcal{H}_{-\bm k}$ at every $\bm k$ is symmetric with respect to $E=0$ (Fig. \ref{fig_1}a).

Due to the combined effect of field-induced Fermi surface displacement 
and Cooper pair momentum, 
the gap structure at $B\neq 0$ differs from the BCS state at $B=0$.  
To the first order of $q$ and $B$,
\begin{eqnarray}\label{eq_e}
{E}_{\pm}(\bm k)=\sqrt{(\xi_{\bm k}\pm |\bm g_{\bm k}|)^2+\Delta^2}+\frac{1}{2}\bm v_{\bm k}\cdot\bm q \pm \hat{\bm g}_{\bm k}\cdot\bm B,
\end{eqnarray}
where $\bm v_{\bm k}=\partial_{\bm k}\xi_{\bm k}|_{\xi=0}=v_{\rm F}\hat{\bm k}$ is the electron velocity and $\hat{\bm g}_{\bm k}=\bm g_{\bm k}/|\bm g_{\bm k}|$. 

{Note that in the Rashba case we have $\hat{\bm g}_{\bm k}\cdot\bm B=-\bm v_{\bm k}\cdot\bm k_0$ and $\bm q_0=-2\bm k_0$ so that $\frac{1}{2}\bm v_{\bm k}\cdot\bm q_0-\hat{\bm g}_{\bm k}\cdot\bm B=0$.}
As shown in Fig. \ref{fig_1}, at small $B$ the outer Fermi surface has a constant superconducting gap $\Delta^{-}({\bm k})=\Delta$ {unaffected by the Zeeman field}, while the inner Fermi surface is affected by the pair-breaking effect of the magnetic field and exhibits a strongly direction dependent gap:  
$
\Delta^{+}({\bm k})= \Delta +\bm v_{\bm k}\cdot\bm q_0.
$
In the case of Rashba superconductors, ${\bm q}_0 \perp \bm B$ and the superconducting gap is most strongly reduced in the direction perpendicular to the field.

At low temperatures, passing a supercurrent through the system creates an {\it additional} phase gradient 
in the superconducting order parameter, so that the current-carrying state has a Cooper pair momentum $\bm q$ 
different from the equilibrium one ${\bm q}_0$. 
When the current is along the axis of ${\bm q}_0$,  
the change of Cooper pair momentum 
$\delta\bm {q} \equiv\bm q -\bm q_0$ is roughly proportional to the supercurrent density $\bm J$ 
and further reduces the superconducting gap. 
As $|\bm J|$ increases, the gap eventually closes. This gap-closing condition provides a rough estimate of the critical current.   
In the case of helical superconductors considered here, 
the gap closes on the inner pocket when the current is in the same direction as ${\bm q}_0$ (Fig. \ref{fig_1}c), 
and on the outer pocket when in the opposite direction (Fig. \ref{fig_1}d). 
Since the two pockets have different gaps, the critical currents in opposite directions are different, resulting in supercurrent diode effect.

{Our discussion  so far of the supercurrent diode effect based on Bogouliubov band structures is heuristic.}
{To {rigorously demonstrate} the supercurrent diode effect, one needs to work out the stable phase of superconductivity, which involves five energy scales: the bare pairing potential $\Delta_0$ at zero temperature and zero field, the spin-orbit coupling energy $\Delta_{so}=\alpha_{\rm R}k_{\rm F}$, the Zeeman energy due to the magnetic field $B$, the temperature $T$ and Fermi energy $\mu$. To determine the complete phase diagram in the five-dimensional parameter space is enormously difficult. Nonetheless, near superconducting-normal phase transitions Ginzburg-Landau theory applies, which may shed light on understanding the whole phase diagram.}

In the following we will present a theory of supercurrent diode effect. We first derive the explicit form of Ginzburg-Landau free energy from symmetry arguments, then calculate the nonreciprocal critical current from such free energy.\\

\begin{figure}
\begin{center}
\leavevmode\includegraphics[width=1\hsize]{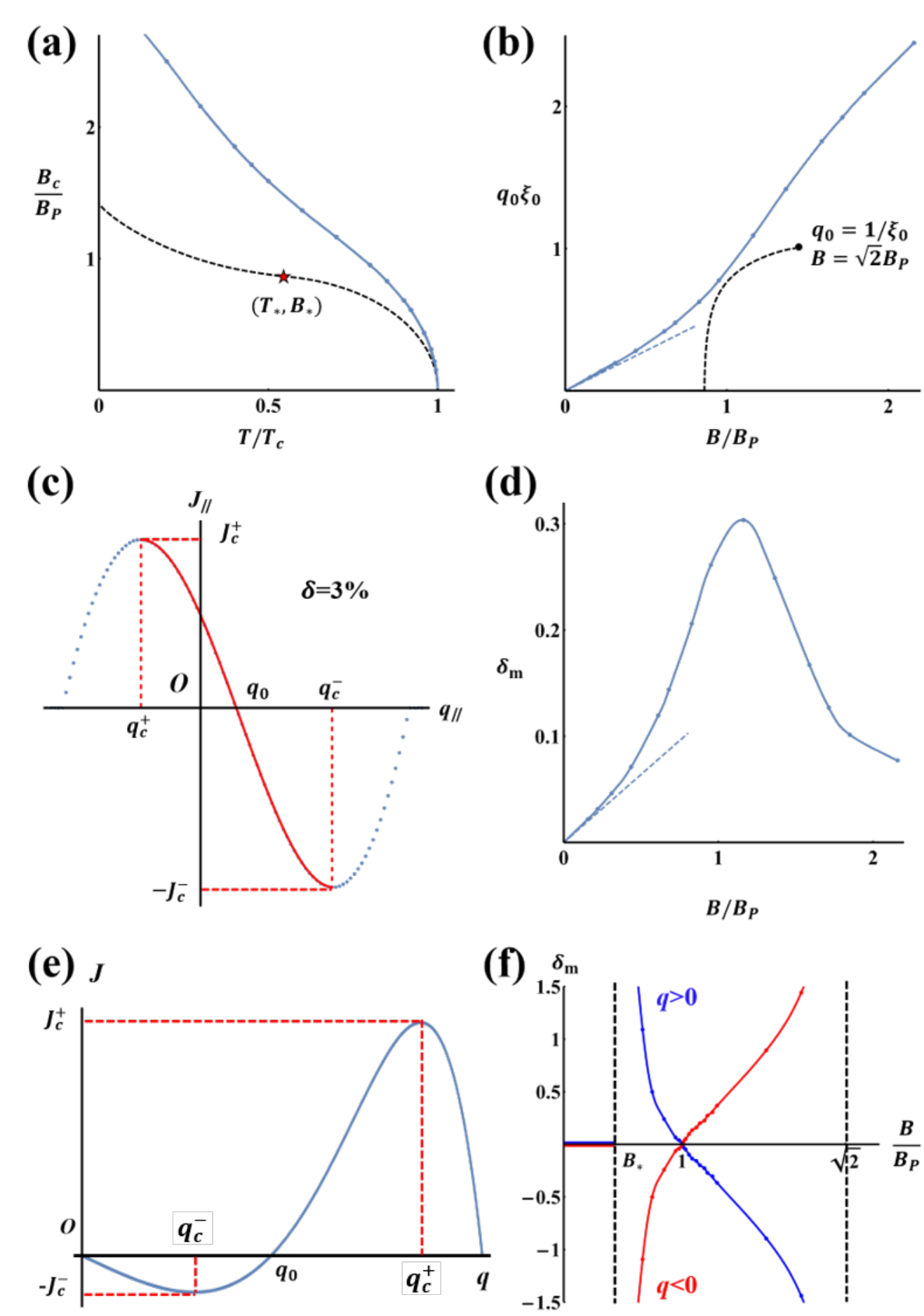}
\end{center}
\caption{(a) In-plane critical field $B_c(T)$ as a function of temperature, for superconductors without SOC (dashed black line) and with Rashba SOC (blue line), where $T_c$ is zero-field critical temperature and $B_P=1.25T_c$ is the Pauli limit. In the BCS case, the red star denotes the tricritical point of FFLO transition. In the Rashba superconductor, $\alpha_{\rm R}=0.2v_{\rm F},\Delta_{so}=2\Delta_0$ and $\Delta_0=1.76T_c$ is zero-temperature order parameter at zero field.
(b) Along the curve $B_c(T)$ in (a), field dependence of Cooper pair momentum magnitude $q_0$, where $\xi_0=v_{\rm F}/\Delta_0$ is the zero-temperature coherence length. Dashed {blue} line denotes Eq. (\ref{eq_q0}) at weak fields. {Dashed black line denotes the BCS case, whose maximum Cooper pair momentum is $q_0=\xi_0^{-1}$ when $B=\sqrt{2}B_P$.} (c) Supercurrent $\bm J=J_{\parallel}\hat{\bm q}_0$ as a function of $\bm q=q_{\parallel}\hat{\bm q}_0$ at $T=0.90T_c,B=0.31B_P$. Under external current, red solid line denotes stable states, and blue dots denote unstable states. (d) Along the curve $B_c(T)$ in (a), field dependence of $\delta_{\rm m}$ defined in Eq. (\ref{eq_f2}). Dashed {blue} line denotes Eq. (\ref{eq_f1}) at weak fields. {(e) $J$ versus $q$ in a superconductor without SOC in the FF state ($q>0$). (f) Along the dashed black curve $B_c(T)$ in (a), $\delta_{\rm m}$ as a function of $B$ in a superconductor without SOC. When $B<B_*$, $\delta=0$ and near $B=B_P$ there is a sign change of $\delta$. Blue and red colors denote two types of FF states ($q>0$ and $q<0$).}}
\label{fig_2}
\end{figure}

\textbf{Theory of supercurrent diode effect}--- {We first review the supercurrent in a conventional BCS superconductor. Close to the superconducting phase transition, the free energy density $f \equiv \tilde{f} N_0$ as a functional of the superconducting order parameter $\Delta(\bm r)$ reads
\begin{equation}
    \tilde{f} = t|\Delta(\bm r)|^2 +a_0|(-i\partial_{\bm r}-2e\bm A)\Delta(\bm r)|^2 +\frac{1}{2}\beta |\Delta(\bm r)|^4,
\end{equation}
and the supercurrent density is then
\begin{equation}
\bm J=-\frac{\partial f}{\partial\bm A}=4e a_0 N_0\left\{ {\rm Im}[\Delta^*(\bm r)\partial_{\bm r}\Delta(\bm r)] -2e|\Delta(\bm r)|^2\bm A\right\}.
\end{equation}
where $t\equiv (T-T_c)/T_c$ is reduced temperature, and $a_0,\beta >0$.
When restricted to single-$\bm q$ order parameter $\Delta(\bm r)=\Delta e^{i\bm q\cdot\bm r}$ and zero vector potential $\bm A=\bm 0$, the corresponding free energy density and supercurrent density take the following forms
\begin{eqnarray}
\tilde{f}(\bm q,\Delta)&=&\alpha_{\bm q}|\Delta|^2 +\frac{1}{2}\beta |\Delta|^4,\nonumber \\
\bm J(\bm q)&=&4e N_0 a_0|\Delta|^2\bm q,
\end{eqnarray}
where $\alpha_{\bm q}=t +a_0 q^2$. Notice that $\bm J=2e\partial_{\bm q}f$ holds, and $\partial_{\bm q}\alpha_{\bm q}=2a_0\bm q$ represents the superfluid velocity.
}

For 2D noncentrosymmetric superconductors under a parallel magnetic field, additional terms involving odd-power derivatives of $\Delta(\bm r)$ are allowed in the free energy expression \cite{YuanPNAS}: 
\begin{eqnarray}\label{eq_f}
\tilde{f}=\Delta^*(\bm r)\hat{\alpha}\Delta(\bm r) +\frac{1}{2}\beta |\Delta(\bm r)|^4,
\end{eqnarray}
where $\hat{\alpha}$ is a differential operator involving the spatial derivative $\partial_{\bm r}$. {For example,  $\hat{\alpha}=t -a_0\partial_{\bm r}^2$ in the conventional BCS case.}
Its Fourier transform $\alpha_{\bm q}$ is obtained from pairing susceptibility of the normal state at wavevector $\bm q$, as shown in the Supplemental Information. 
For spin-orbit-coupled superconductors considered in this work, $\beta>0$ generally holds.  
When $\alpha_{\bm q}<0$, the above free energy is minimized by single-$\bm q$ order parameter: $\Delta(\bm r)=\Delta e^{i\bm q\cdot\bm r}$. The corresponding free energy density takes the form
\begin{eqnarray}
\tilde{f}(\bm q,\Delta)=\alpha_{\bm q}|\Delta|^2 +\frac{1}{2}\beta |\Delta|^4.
\end{eqnarray}
The Cooper pair momentum $\bm q_0$ in the equilibrium state is determined by 
minimizing $\alpha_{\bm q}$ over $\bm q$, i.e., 
\begin{eqnarray}
\left.\frac{\partial \alpha}{\partial \bm q } \right\vert_{\bm q_0} =0, \; {\rm and} \;  
\left.\det \frac{\partial^2 \alpha}{\partial q_i \partial q_j } \right\vert_{\bm q_0} >0. 
\end{eqnarray}
The in-plane critical field $B_c(T)$ 
is determined from the condition 
\begin{eqnarray}
\min_{\bm q} \alpha_{\bm q}=0 \; {\rm at } \; B=B_c, 
\end{eqnarray}
where $\alpha_{\bm q}$ is temperature- and field-dependent. Equivalently, one also determines the critical temperature $T_c(B)$ at finite field $B$ in the same way.

As shown above, the single-$\bm q$ order parameter is the stable phase near the superconducting phase transition. However, deep in the superconducting phase, either $B\ll B_c(T)$ or $T\ll T_c(B)$, the multiple-$\bm q$ order parameter (e.g. LO phase) will compete with the single-$\bm q$ one, and might be the stable phase in some cases \cite{Olga2,Wang}. Nevertheless, throughout this work we focus on states close to the critical field line $B=B_c(T)$ whose stable phase is described by a single-$\bm q$ order parameter.

For the Rashba superconductor, 
the in-plane critical field $B_c(T)$ and the corresponding Cooper pair momentum ${\bm q}_0$ 
at the superconducting transition is calculated numerically and shown in Fig. \ref{fig_2}a and \ref{fig_2}b. 
Note that ${\bm q}_0$ increases smoothly from zero with $B$, leading to 
a helical superconductor. The critical field is higher than the case of BCS
phase (dashed line in Fig. \ref{fig_2}a) in the absence of SOC.

By introducing vector potential $\bm A$, the uniform supercurrent density $\bm J$ can be calculated from the free energy density $f$ as follows
\begin{eqnarray}
\bm J=-\left.\frac{\partial f(\bm q-2e\bm A,\Delta)}{\partial\bm A}\right|_{\bm A=\bm 0}=2e\partial_{\bm q}f \label{Jq}
\end{eqnarray}
with electron charge $e<0$. The equilibrium state minimizes free energy density $\partial_{\bm q}f=\bm 0$ and hence carries zero current. 
By connecting the system to an external source, one can pass a nonzero supercurrent $\bm J$ through the system.  
Such current-carrying state has a Cooper pair momentum ${\bm q}\neq {\bm q}_0$ that is determined 
by $\bm J$ according to Eq. (\ref{Jq}). 
Minimizing the free energy with respect to the gap magnitude $\Delta$ 
yields $|\Delta|^2=-\alpha_{\bm q}/\beta$ when $\alpha_{\bm q}<0$. 
Then, Eq. (\ref{Jq}) becomes
\begin{eqnarray}\label{eq_f0}
\bm J=
\frac{eN_0}{\beta}|\alpha_{\bm q}|\partial_{\bm q}\alpha_{\bm q}. 
\end{eqnarray}
Clearly, the supercurrent is the product of the superfluid stiffness $|\alpha_{\bm q}|$ 
and the superfluid velocity $\partial_{\bm q}\alpha_{\bm q}$, both of which depend on the 
Cooper pair momentum $\bm q$. 

We first consider the BCS case without SOC and obtain the superconducting phase diagram in $B$-$J$ plane 
for temperatures close to the zero-field and zero-current critical temperature $T_c$. 
At $B=0$, $\alpha_{\bm q}=t+a_0 q^2=t(1-\xi^2 q^2)$ with coherence length $\xi\equiv\sqrt{a_0/|t|}\propto(T_c-T)^{-\frac{1}{2}}$, 
so that ${\bm J} \propto (1-\xi^2 q^2) \xi^2 {\bm q} $. 
At $q=0$, the superfluid stiffness is maximal 
but the superfluid velocity vanishes. 
At $q=\xi^{-1}$, the velocity is high but the stiffness vanishes. 
The maximal supercurrent or the critical current $J_c$ is achieved at an intermediate momentum 
$q_c=\xi^{-1}/\sqrt{3}$ with $J_c\propto(T_c-T)^{3/2}$ \cite{Aarts}.
At $B\neq 0$ without external current, the Zeeman effect of in-plane magnetic field  
decreases the critical temperature of a BCS superconductor, 
leading to $T_c-T_c(B)\propto B^2$ or equivalently $B_c(T) \propto ({T_c - T})^{1/2}$.

When both $B$ and $J$ are nonzero, 
the boundary between superconducting phase and normal phase in $B$-$J$ plane is defined by 
\begin{eqnarray}\label{BCS}
\left(\frac{B}{B_c}\right)^2+\left(\frac{J}{J_c}\right)^{2/3}=1,
\end{eqnarray}
where exponents $2$ and $2/3$ follow from the temperature scaling of the zero-current critical field $B_c$ 
and the zero-field critical current $J_c$. The phase boundary is in general smooth except the non-analytical region near $J=0$ due to the fractional exponent 2/3.
This phase boundary is symmetric in $J$ 
due to inversion symmetry of a conventional superconductor 
$\bm B\to\bm B,\bm J\to -\bm J$, and also symmetric with respect to the origin $J=B=0$ due to 
time-reversal symmetry $\bm B\to -\bm B,\bm J\to -\bm J$.

In superconductors without inversion symmetry, 
the phase boundary in $B$-$J$ phase space can become skewed as sketched in Fig. \ref{fig_0}d, 
in which case two closely-related nonreciprocal effects appear. 
At a given magnetic field $B\neq 0$, the critical currents are nonreciprocal; 
and in a given current-carrying state $J\neq 0$, the critical fields are polarity dependent. 
Both phenomena are manifestations of the supercurrent diode effect \cite{Yuta,TeruoNature}.

As a concrete example, we consider Rashba superconductors in the following and derive explicitly the nonreciprocal critical current at weak field and temperatures near $T_c$.
First, at small $B$ we can expand $\alpha_{\bm q}$ as a power series in $\bm q$ and keep terms up to linear order in $B$ \cite{YuanPNAS},
\begin{eqnarray}\label{eq_alpha}
\alpha_{\bm q}=t-(b_0 -b_1 q^2)\bm q\cdot ({\bm B\times\hat{\bm z}})+a_0q^2 , \label{alphaq}
\end{eqnarray}
with the parameters $a_{0},b_{0},b_1$ derived from Fermi surface integrals as detailed in the Supplemental Information:
\begin{eqnarray}
b_0=C_0\frac{\alpha_{\rm R}}{(\pi T_c)^2},\; b_1=C_1\frac{v_{\rm F}^2\alpha_{\rm R}}{(\pi T_c)^4}, \nonumber \;
a_0=\frac{1}{4}C_0\frac{v_{\rm F}^2}{(\pi T_c)^2}, 
\label{parameters}
\end{eqnarray}
where $C_0=1.04$, $C_1=0.38$ are numerical constants. In deriving the expressions of $a_{0},b_{0}, b_1$ we have assumed spin-orbit coupling strength $\Delta_{so}$ is larger than the superconducting gap $\Delta_0$. 
In Eq. (\ref{eq_alpha}), 
the $B$-linear terms arise from the field-induced shift ($b_0$) and deformation ($b_1$) of Fermi surfaces. 
By minimizing $\alpha_{\bm q}$ over $\bm q$, we find the   
Cooper pair momentum in the equilibrium state 
\begin{eqnarray}\label{eq_q0}
{\bm q}_0 = 2\alpha_{\rm R}(\bm B\times\hat{\bm z})/v^2_{\rm F}
\end{eqnarray} to the leading order in $B$. 

Substituting Eqs. (\ref{alphaq}) and (\ref{parameters}) into Eq. (\ref{eq_f0}), 
we obtain the supercurrent along the axis parallel to $\bm q_0$ direction $J_{\parallel}$ as a function of $q_{\parallel}$, 
shown in Fig. \ref{fig_2}c.
Due to the third-order term, $\alpha_{\bm q}$ is skewed 
with respect to its minimum at ${\bm q}_0$.
This skewness leads to different critical currents in the direction parallel and antiparallel to ${\bm q}_0$.
More specifically, upon the application of a weak in-plane magnetic field, critical current increases along one direction while decreases along the opposite direction
\begin{equation}\label{eq_f1}
    \frac{J_{c}^{\pm}}{J_{c}}=1\pm \gamma\frac{B}{B_c} 
\end{equation}
with 
\begin{eqnarray}\label{delta0}
\gamma(T)= {0.64}\frac{\alpha_{\rm R}}{v_{\rm F}}\frac{B_c}{B_P}\sqrt{1-\frac{T}{T_c}}
\end{eqnarray}
where $B_P=1.25T_c$ is the Pauli limit. The dimensionless quantity $\gamma$ measures the strength of field-induced supercurrent diode effect at temperature $T$. 
$\gamma(T)$ is proportional to Rashba SOC and decreases to zero as $T_c - T$ near the critical temperature. 
Since $J_c\propto(T_c-T)^{3/2}$ and $B_c\propto(T_c-T)^{1/2}$, we have $J_c^+-J_c^-\propto(T_c-T)^2$. 

More generally, at higher fields the Cooper pair momentum $q_0$ is no longer small and reaches the order of $\xi_0^{-1}$ at $B\sim B_P$ (see Fig. \ref{fig_2}b), where $\xi_0$ is the zero-temperature, zero-field and zero-current coherence length. Nonetheless, the critical current is always small near the superconducting transition temperature $T_c(B)$ at the corresponding field. Under this condition,
we can expand $\alpha_{\bm q}$ around its minimum $\bm q_0$:  
\begin{eqnarray}\label{eq_alpha1}
\alpha_{\bm q}=A[T-T_c(B)]+a\delta q_{\parallel}^2-b\delta q_{\parallel}^3,
\end{eqnarray}
where $A,a>0$ and $\delta q_{\parallel}=(\bm q-\bm q_0)\cdot\hat{\bm q}_0$. From this expression we find different critical currents $J_c^\pm$ in the direction along and against ${\bm q}_0$. The nonreciprocity of critical current can be characterized by a ``supercurrent diode coefficient'' which we define as  
\begin{eqnarray}
  \delta \equiv \frac{J_c^+ - J_c^-}{J_c^+ + J_c^-}.
\end{eqnarray}
At temperatures close to the superconducting phase transition, we find 
\begin{equation}\label{eq_f2}
  \delta =\delta_{\rm m}\sqrt{1-\frac{T}{T_c(B)}}, \;{\rm with } \; \delta_{\rm m}=b\sqrt{\frac{AT_c(B)}{3a^3}}.
\end{equation}

Our theory based on Eq. (\ref{eq_alpha1}) applies to both the regime $T\sim T_c$ at weak fields and also $T\sim 0$ at high fields since $T_c(B)\to 0$ when $B$ is large (Fig. \ref{fig_2}a). The coefficients $a,b$ can be computed by derivatives $\partial^2_q\alpha, \partial^3_q\alpha$ at $\bm q=\bm q_0$. 
One can numerically compute $a$ and $b$ from $\alpha_{\bm q}$ as shown in Supplemental Information, and $\delta$ can be obtained as shown in Fig. \ref{fig_2}c and d.

We numerically calculate and plot $\delta_{\rm m}$ along the critical field line $B=B_c(T)$, as shown in Fig. \ref{fig_2}d. Detailed derivation of Eqs. (\ref{eq_f1},\ref{delta0},\ref{eq_alpha1}) and (\ref{eq_f2}) can be found in Supplemental Information.

It can be seen that $\delta_{\rm m}$ is linear in $B$ at weak fields as expected in Eq. (\ref{eq_f1}) where $\delta=\gamma B/B_c$. As field increases, $\delta_{\rm m}$ reaches its maximum near $B=B_{P}$, and then decreases again.
To understand the behavior of $\delta_{\rm m}$ under magnetic field, note that $\alpha_{\bm q}=1/V-\chi_{\bm q}$  ($V$ is the attractive interaction) is determined by pairing susceptibility $\chi_{\bm q}=\chi^{+}_{\bm q}+\chi^{-}_{\bm q}$, where $\chi_{\bm q}^{\pm}$ are the contribution from inner (+) and outer ($-$) Fermi surfaces respectively.
$\chi_{\bm q}^{\pm}$ are nearly symmetric with respect to their respective maximum at $\bm q \sim \pm 2 {\bm k}_0$ due to the Fermi surface shift, with $k_0$ on the order of $B/v_{\rm F}$. The peak width is on the order of $T/v_{\rm F}$. Due to the DOS asymmetry, the peak value of $\chi^{-}_{\bm q}$ is higher than that of $\chi^{+}_{\bm q}$. The differences in the peak position and height of $\chi^{+}$ and $\chi^{-}$  give rise to the skewness of $\chi_{\bm q}$, the sum of the two.
At high temperature $T\lesssim T_c$ and low field $B\ll B_P$, the peaks of $\chi_{\bm q}^{\pm}$ are relatively broad and close to each other, so that the asymmetry in $\alpha_{\bm q}$ is small. 
In the opposite limit of low temperature $T\ll T_c$ and high field $B\sim B_c(T)\gg B_P$, the two peaks are narrow and well separated and differs greatly in the height. $\chi_{\bm q}$ is dominated by the main peak $\chi^{-}_{\bm q}$ and therefore is nearly symmetric.
Since the nonreciprocity in critical current becomes small at both low and high fields, the supercurrent diode coefficient $\delta_{\rm m}$---a measure of nonreciprocity near $T_c(B)$---reaches its maximum around the Pauli limit $B=B_{P}$.
We also discuss in the Supplemental Information the possibility of nonreciprocal critical current in the single-$\bm q$ FF (helical) phase without SOC.  

Next, we derive the polarity-dependent critical field at small current for temperatures near $T_c$. 
This requires including the correction to $\alpha_{\bm q}$ at second order in $B$ as in usual BCS superconductors without SOC.  
As shown in the Supplemental Information, we obtain the skewed phase boundary for Rashba superconductors in $B-J$ plane: 
\begin{eqnarray}\label{Smectic}
\left(\frac{B}{B_c}\right)^2+\left|\frac{\bm J}{J_c}-\gamma\frac{\bm B\times\hat{\bm z}}{B_c}\left(1-\frac{B^2}{B^2_c}\right)^2\right|^{2/3}=1,
\end{eqnarray}
where $\gamma$ is defined in Eq. (\ref{delta0}). 
To the leading order in $B$, we recover Eq. (\ref{eq_f1}) for nonreciprocal critical currents, and to the leading order in $J$, we find the polarity-dependent critical field
\begin{eqnarray}\label{eq_f3}
\frac{B_{c}^{+}-B_{c}^{-}}{2B_c}=\frac{\gamma}{3}\frac{J}{J_c},
\end{eqnarray}
where the factor $1/3$ is from exponents 2 and 2/3.
Since $J_c\propto(T_c-T)^{3/2}$, $B_c\propto(T_c-T)^{1/2}$ and $\gamma\propto(T_c-T)$, we find $B_c^+-B_c^-$ is temperature-independent near $T_c$. Eq. (\ref{Smectic}) is our key finding about the supercurrent diode effect: it is exact (in the sense of temperature scaling) at temperatures near the zero-field,  zero-current $T_c$.

At low temperatures, the entire phase boundary in $B$-$J$ plane can be determined  numerically. 
From Eq. (\ref{Smectic}), we find the supercurrent diode effect in Rashba superconductors is maximized when $\bm B\perp\bm J$, and vanishes when $\bm B\parallel\bm J$, as shown in Fig. \ref{fig_0}b and d. 
The latter property is guaranteed when mirror symmetry $\bm B\to\bm B,\bm J\to -\bm J$ is present. 

Besides Rashba systems, helical phase exists in noncentrosymmetric superconductors with the following point groups: ${\rm D}_{n},{\rm C}_{n\rm v},{\rm C}_{n},{\rm D}_{2\rm d},{\rm S}_{4},{\rm C}_1 (n=2,3,4,6)$. In these systems, crystal symmetry allows a linear coupling between the Cooper pair momentum $\bm q$ and Zeeman field $\bm B$ enabled by spin-orbit coupling. 
As a result, the field-induced Cooper pair momentum ${\bm q}_0$ is linearly proportional to $\bm B$ at weak fields \cite{YuanPNAS} : 
\begin{eqnarray}\label{eq_qB}
\bm q_0\propto\nabla_{{\bm k}}(\hat{\bm g}_{\bm k}\cdot\bm B)|_{\bm k=\bm 0}.  
\end{eqnarray}
The form of spin-orbit vector $\bm g_{\bm k}$ and therefore the direction of Cooper pair momentum 
depend on crystal symmetry. 
While $ \bm q_0\propto \hat{\bm z} \times \bm B$ is perpendicular to the in-plane magnetic field in Rashba superconductors, 
$ \bm q_0\propto\bm B $ is parallel to $\bm B$ in crystals with ${\rm D}_{n \geq 3}$ point groups, and $\bm q_0\propto(B_x,-B_y)$ forms a mirror pair with $\bm B$ in crystals with  ${\rm D}_{\rm 2d}$ point group. While the direction of Cooper pair momentum relative to the magnetic field depends on crystal symmetry, the supercurrent diode effect is generally allowed when the current is passed along the axis of Cooper pair momentum. 

{For Ising superconductors with point group ${\rm D}_{\rm 3h}$ such as transition metal dichalcogenides with an odd number of layers, the supercurrent diode effect is absent for in-plane field and in-plane current, where the combined symmetry of vertical mirror $I_z: z\to -z$ with respect to the basal plane and time-reversal $\mathcal{T}$ is preserved $I_z\mathcal{T}: \bm J\to-\bm J,\bm B\to\bm B$. However, one can realize nonreciprocal critical currents in an Ising superconductor by introducing an out-of-plane magnetization $M_z$ (e.g. by ferromagnetic proximity effect), so that
\begin{equation}
\alpha_{\bm q}=t+a_0 q^2 +b(q_x^3 -3q_x q_y^2)M_z,
\end{equation}
and $a_0>0$, $b$ are determined by Fermi surface properties of the Ising superconductor.
Denoting $\theta$ as the angle between $\bm J$ and $x$-axis, the supercurrent diode coefficient is then
\begin{equation}
\delta=b\sqrt{\frac{|t|}{3a_0^3}} M_z\cos 3\theta.
\end{equation}}

{We further consider supercurrent diode effect in the limit of vanishing SOC. In this case, 
finite momentum FFLO superconductivity occurs at high magnetic fields above the Pauli limit. 
In particular, the single-$q$ FF state (which can be energetically favored over the LO state by weak SOC) breaks both inversion and time reversal symmetry, thus giving rise to the diode effect as shown in Fig.\ref{fig_2}e.    
In Fig. \ref{fig_2}f, we also plot $\delta_{\rm m}$ along the critical field line $B=B_c(T)$, where 
the normal phase, the BCS phase and the FF phase meet at the tricritical point $(T_{*},B_{*})$ (see Fig. \ref{fig_2}a). 
As the magnetic field increase from $B_*$ to $B_c=\Delta=\sqrt{2} B_P$, the Cooper pair momentum increases rapidly from $0$ to $1/\xi_0$, and the diode coefficient $\delta$ shows remarkable features. }

%
{The magnitude of the supercurrent diode effect $|\delta|$ becomes very large near both $B_*$ (the tricritical point) and $B_c$ (the $T=0$ end point of the FF phase). This can be understood from the behavior of GL coefficient $\alpha_{\bm q}$. Near the tricritical point, we have   
\begin{equation}\label{eq_f1cf}
\alpha_{\bm q}={c}_{0} + c_1 q^2+ c_2 q^4,
\end{equation}
with $c_0\propto T-T_*$, $c_1\propto B_*-B$, and $c_2>0$. At $B$ just above $B_*$, the Cooper pair momentum  $q_0\propto \sqrt{B-B_*}$ is small. Nonetheless, $\alpha_q$ is highly skewed with respect to ${\bm q}_0$: it increases very slowly as $q$ decreases from $q_0$. As a result, $J_{c}^{+} \rightarrow 0$ while $J_{c}^{-}$ remains finite as $T\to T_*,B\to B_*$. Thus the diode coefficient  $\delta=(J_c^+-J_c^-)/(J_c^++J_c^-) \sim 1$ reaches the maximum possible value near the tricritical point.} 

{On the other hand, near the transition between FF superconductor and normal phase at low temperature $T\rightarrow 0$, we have  
\begin{equation}\label{eq_f1b}
\alpha_{\bm q}={\rm Re} \left[\log\left(\frac{B+\sqrt{B^2-v_{\rm F}^2q^2}}{\Delta_0}\right) \right].
\end{equation}
Minimizing $\alpha_{\bm q}$ over $\bm q$ yields a large Cooper pair momentum $q_0=B/v_{\rm F}$ approaching $1/\xi_0$ as $B\rightarrow B_c$.  In this case, due to the non-analytic dependence of $\bm q$, $\alpha_{\bm q}$ is highly skewed with respect to ${\bm q}_0$: it rises steeply as $q$ decreases from $q_0$. This leads to the maximum possible  diode effect with $\delta=(J_c^+-J_c^-)/(J_c^++J_c^-) \sim -1$ near $B_c$, taking opposite sign as the one near tricritical point.
Details of the supercurrent diode effect in the FF phase can be found in the Supplemental Information. Note that at low temperature away from the critical line $B_c(T)$,  single-$\bm q$ FF state may yield to multiple-$\bm q$ state. In the absence of SOC, the double-$\bm q$ LO state has a spatially modulated gap amplitude and constant phase, which does not show the diode effect.    
}


\textbf{Discussion}

The supercurrent diode effect predicted in this work,  including nonreciprocal critical current and polarity-dependent in-plane critical field, may be observed in polar SrTiO$_3$ films \cite{STO} and artificially engineered heavy fermion superlattices of
YbCoIn$_5$-YbRhIn$_5$-CeCoIn$_5$ \cite{Matsuda}, where the layer stacking breaks inversion symmetry.
In both cases, an upturn of in-plane critical field has been observed and attributed to 
the Rashba spin splitting. Other candidates include transition metal dichacodegnide MoTe$_2$ in the low-symmetry $T_d$ structure \cite{MoTe2,MoTe2-Dan,MoTe2-Law}, and half-heusler superconductors such as YPtBi with tetrahedral point group \cite{YPtBi1,YPtBi2,YPtBi3,YPtBi4}. 
These two classes of materials have strong spin-orbit interaction and small Fermi energy, which
enhances the supercurrent diode effect.  

Another interesting platform to search for supercurrent diode effect is  
magic-angle twisted bilayer graphene \cite{TBG1,TBG2}, whose twisted bilayer structure breaks inversion symmetry. 
Although spin-orbit interaction is negligibly small, an in-plane magnetic field modifies   
the moir\'e band structure by giving electrons a momentum shift in interlayer tunneling process. 
As we showed recently, a small field causes a shift of Fermi surface by ${\delta \bm k} \propto \bm B$ 
in the direction {\it parallel} to $\bm B$ \cite{TBG2}. 
Therefore field-induced finite-momentum superconductivity 
and supercurrent diode effect may appear when the critical current is measured along the field axis.

Last but not the least, the supercurrent diode effect should also exist when the superconducting state spontaneously breaks time-reversal and  inversion symmetry. Thus  nonreciprocal critical current at zero magnetic field and  polarity-dependent critical field in the equilibrium state provide new probes of unconventional superconductivity with hidden orders.

\textbf{Data Availability.} All study data are included in this article.

\textbf{Acknowledgments.}
We thank Yang Zhang for bringing Ref. \cite{TeruoNature} to our attention. 
This work is supported by DOE Office of Basic Energy Sciences, Division of Materials Sciences and Engineering under Award DE-SC0018945. 
L.F. is partly supported by a Simons Investigator award from the Simons Foundation.

\textit{Note Added}---
Recently two related works \cite{Nagaosa3,Yanase} on the similar topic have appeared.

\newpage

\appendix
\setcounter{figure}{0}
\renewcommand{\thefigure}{S\arabic{figure}}
\setcounter{equation}{0}
\renewcommand{\theequation}{S\arabic{equation}}

\section{Ginzburg-Landau coefficient}
We compute the the function $\alpha_{\bm q}$ microscopically within mean field theory. From Eqs. (1,4) of the main text, the mean-field free energy density is
$f(\bm q,\Delta)=|\Delta|^2/V-T\int d^2\bm k{\rm tr}\log[1+e^{-\mathcal{H}_{\bm k}(\bm q,\Delta)/T}]$, where $V$ is the attractive interaction strength and tr denotes trace in spin space. 

By expansion of free energy density $f(\bm q,\Delta)$ in terms of $\Delta$ we have
\begin{widetext}
\begin{eqnarray}\label{eq_f1cs}
\alpha =\log\frac{T}{T_c}+\frac{1}{N_0}\sum_{\lambda=\pm}\oint_{{\rm FS}_{\lambda}}\frac{dk}{|\bm v^\lambda|} \left\{\phi \left(\frac{Q+\lambda\mathcal{E}_{+}}{2\pi T}\right)\cos^2\frac{\theta}{2}
+ \phi \left(\frac{Q+\lambda\mathcal{E}_{-}}{2\pi T}\right)\sin^2\frac{\theta}{2}\right\},
\end{eqnarray}
\end{widetext}
where $\lambda=\pm$ denote contributions from inner ($\lambda=+$) and outer ($\lambda=-$) Fermi surfaces respectively,
$\phi(x)={\rm Re}\left[\psi\left(\frac{1+ix}{2}\right)\right]-\psi\left(\frac{1}{2}\right),$
and $\psi$ is the digamma function. 

Here,  $\bm v^\lambda=\partial_{\bm k}\xi_{\bm k}^{\lambda}$ is the electron velocity,
$Q=\bm v^{\lambda}\cdot\bm q$ is the depairing energy of finite momentum pairing, $ \mathcal{E}_{\pm}=|\bm h_{+}|\pm |\bm h_{-}| $ is the depairing energy of Zeeman splitting for inter- ($+$) or intra-pocket ($-$) Cooper pairs with $\bm h_{\pm}=\bm B+\bm g_{\frac{1}{2}\bm q\pm\bm k}$, and the angle $ \theta=\langle\bm h_{+},\bm h_{-}\rangle $ between $\bm h_{\pm}$ controls the ratio between inter- or intra-pocket Cooper pairs.
Supercurrent affects $\bm q$ and hence depairing energy $Q$, while magnetic field $\bm B$ together with SOC affects depairing energies $\mathcal{E}_{\pm}$ and angle $\theta$. 

Near $T_c$, the temperature dependence of $\alpha_{\bm q}$ can be captured by the first term $\log(T/T_c)$, and we can set $T=T_c$ in the Fermi surface integrals. To evaluate the integral, notice that the field is weak $B\ll B_P$, one can expand the special function 
\begin{eqnarray}
\phi(x)=2.10x^2-2.01x^4+2.00x^6+O(x^8)
\end{eqnarray}
and then integrate order by order to obtain Eq. (14) of the main text.

\section{Nonreciprocal critical current and polarity-dependent critical field}
For $\alpha_{\bm q}=\overline{\alpha}+a\delta q_{\parallel}^2-b\delta q_{\parallel}^3,$ we find the 
supercurrent is
\begin{eqnarray}
\beta J_{\parallel}/e=|\alpha_{\bm q}|\partial_{\parallel}\alpha_{\bm q}=2 a \overline{\alpha} \delta q_{\parallel} -3 \left(\overline{\alpha} b\right)\delta q_{\parallel}^2\\
+2 a^2 \delta q_{\parallel}^3 - 5 (a b) \delta q_{\parallel}^4 +3b^2\delta q_{\parallel}^5.
\end{eqnarray}
Notice that to the leading order of $|\overline{\alpha}|^{\frac{1}{2}}$, critical currents $\pm J_{c}^{\pm}$ correspond to $\delta q_{\parallel}=\mp \delta q_c$ respectively, where $\delta q_c=\sqrt{|\overline{\alpha}|/3a}$. Then we have
\begin{eqnarray}\label{eq_jc}
\beta J_{c}^{\pm}/e=\frac{4 |\overline{\alpha}|^{3/2}}{9 a}\left(\sqrt{3a^3}\pm b\sqrt{|\overline{\alpha}|}\right) +O(|\overline{\alpha}|^{5/2}).
\end{eqnarray}

Since $\overline{\alpha},a,b$ are all functions of $B$ and $T$, the equation above determines two phase boundaries parametrized by $J$, $B$ and $T$ as depicted in Fig.  1b of the main text. 
At weak field, $a$ can be treated as a constant, $b\propto B$ and $\overline{\alpha}= t(1-B^2/B^2_c)$ with reduced temperature $t=(T-T_c)/T_c$ and critical field $B_c\propto|t|^{1/2}$. Then Eqs. (16-22) of the main text can be obtained.

Especially for Rashba superconductors at weak field, from Eq. (13) of the main text and (\ref{eq_jc}) we obtain 
\begin{eqnarray}
\frac{\beta\bm J}{eN_0}=\frac{4}{3\sqrt{3}}|{\alpha}_{\bm q_0}|^{2}\xi\hat{\bm n}+\frac{4}{9}\frac{b_1|{\alpha}_{\bm q_0}|}{\xi^2}\bm B\times\hat{\bm z},
\end{eqnarray}
where $\hat{\bm n}=(\bm q-\bm q_0)/|\bm q-\bm q_0|$, and under magnetic field
\begin{equation}
\alpha_{\bm q_0}= t\left(1-\frac{B^2}{B^2_c}\right),\;  \xi= C \frac{v_{\rm F}}{\pi T_c}\left|t\left(1-\frac{B^2}{B^2_c}\right)\right|^{-\frac{1}{2}}.
\end{equation}
Then we obtain the skewed phase boundary
\begin{eqnarray}
\left(\frac{B}{B_c}\right)^2+\left|\frac{\bm J}{J_c}-\gamma\frac{\bm B\times\hat{\bm z}}{B_c}\left(1-\frac{B^2}{B^2_c}\right)^2\right|^{2/3}=1,
\end{eqnarray}
with zero-field critical current
\begin{eqnarray}
J_c=\frac{4eN_0^2C}{3\sqrt{3}\beta} \frac{v_{\rm F}}{\pi T_c}|t|^{3/2}
\end{eqnarray}
and the skewness parameter
\begin{eqnarray}
\gamma =\frac{b_1 B_c |t|^{1/2}}{\sqrt{3}N_0}\left(\frac{\pi T_c}{v_{\rm F}}\right)^{3}=0.64\frac{\alpha_{\rm R}}{v_{\rm F}}\frac{B_c}{B_P}\sqrt{1-\frac{T}{T_c}}.
\end{eqnarray}

To include higher order contributions, the supercurrent diode coefficient is
\begin{equation}
    \delta=D\left(x\right),\quad x=\sqrt{\frac{|\alpha_{\bm q_0}|}{a^3}} {b}.
\end{equation}
The special function is
\begin{eqnarray}
D(x)=\frac{J(x)-J(-x)}{J(x)+J(-x)},
\end{eqnarray}
where $J(x)=[Q(x)-\frac{3}{2}xQ(x)][-1+Q(x)^2-xQ(x)^3]$,
\begin{equation*}
Q(x)=\frac{1}{3 x}-\frac{\sqrt{n}}{2 x}-\frac{1}{2} \sqrt{{\frac{4}{5\sqrt{n}}}\left(1-\frac{2}{27 x^2}\right)-\frac{n}{x^2}+\frac{8}{15 x^2}}
\end{equation*}
and $n=\frac{2}{15}\left(z+1/{z}+\frac{4}{3}\right),$
\begin{eqnarray*}
z=\sqrt[3]{t+\sqrt{t^2-1}},\quad
t=\frac{135}{4}x^4-5 x^2+1. 
\end{eqnarray*}
When $|x|<\frac{2}{3\sqrt{3}}$, $\frac{22}{27}<t<1$ and $z$ is complex. Denote $t=\cos\theta$, then $z=e^{i\theta/3}$ and $n=\frac{4}{15}(\cos\frac{\theta}{3}+\frac{2}{3})$ is real.
Since $D(x)\approx x/\sqrt{3}$ we get the leading order contribution $\delta=\sqrt{\frac{|\alpha_{\bm q_0}|}{3a^3}} {b}$, namely
{
\begin{equation}
  \delta =\delta_{\rm m}\sqrt{1-\frac{T}{T_c(B)}}, \;{\rm with } \; \delta_{\rm m}=b\sqrt{\frac{a_0T_c(B)}{3a^3}}.
\end{equation}}

{
The expansion of $\alpha_{\bm q}$ near its minimum $\bm q_0$ in general can be anisotropic
\begin{eqnarray}
\alpha_{\bm q+\bm q_0}=\alpha_{\bm q_0}+a(1+\epsilon) q_x^2+a(1-\epsilon)q_y^2 +2a\eta q_x q_y \\\nonumber
-(b_1 q_x^3 +b_2 q_y^3 +b_3 q_x q_y^2+b_4 q_x^2 q_y),
\end{eqnarray}
where $a>0$ and $\epsilon^2+\eta^2<1$ for stability. The supercurrent diode coefficient for supercurrent $\bm J=J(\cos\theta,\sin\theta)$ can be worked out as
\begin{widetext}
\begin{eqnarray}\label{eq_diode}
\delta =\sqrt{\frac{|\alpha_{\bm q_0}|}{3a^3}}\frac{\left(\frac{b_1+b_3}{2}+\frac{b_1-b_3}{2}\cos2\theta\right)\cos\theta+\left(\frac{b_2+b_4}{2}-\frac{b_2-b_4}{2}\cos2\theta\right)\sin\theta}{(1+\epsilon\cos 2\theta+\eta\sin 2\theta)^{3/2}}.
\end{eqnarray}
\end{widetext}}

\section{Supercurrent diode effect near FFLO transition}
When magnetic field is high, near a phase transition where two or more local minima of $\alpha_{\bm q}$ compete, strong supercurrent diode effect can happen. 
As a concrete example, we consider the transition from BCS phase to FF phase near the upturning point $(T_{*},B_{*})$ (red star in Fig. 3a of the main text), and adapt the following free energy density expanded up to quartic order in $q$
\begin{equation}\label{eq_f1bs}
\alpha_{\bm q}={c}_{0} + c_1 q^2+ c_2 q^4,
\end{equation}
where $c_1=c(B_*^2-B^2)$, and $c,c_2>0$. 
When $B<B_*$, $c_1>0$ and the BCS phase with zero Cooper pair momentum is the ground state. When $B>B_*$, $c_1<0$ and the ground state changes to the FF phase with Cooper pair momentum $q_{0}\equiv\sqrt{|c_1|/(2c_2)}\propto\sqrt{B^2-B_*^2}$. As a result, when one lowers the temperature, the in-plane critical field $B_{c}(T)$ exhibits an upturn across the upturning point and hence the name. 

To better understand the FFLO physics we need quartic order Ginzburg-Landau analysis. One can write the order parameter in Fourier form $\Delta(\bm r)=\int d^2\bm q\Delta_{\bm q}e^{i\bm q\cdot\bm r}$ and the free energy then reads
\begin{widetext}
\begin{eqnarray}\label{eq_4GL}
F\equiv \int d^2\bm r f(\bm r)=\int d^2\bm q\alpha_{\bm q}|\Delta_{\bm q}|^2 +\frac{1}{2}\int d^2\bm q d^2\bm p d^2\bm p'\beta_{\bm q\bm p\bm p'} \Delta_{\bm q+\bm p} \Delta_{\bm q-\bm p}\Delta^*_{\bm q+\bm p'} \Delta^*_{\bm q-\bm p'}.
\end{eqnarray}
\end{widetext}

By analyzing the quartic coefficients $\beta_{\bm q\bm p\bm p'}$ and the quadratic coefficients $\alpha_{\bm q}$ one can then distinguish FF and LO phases \cite{Olga2,Wang}. The quartic coefficient $\beta_{\bm q\bm p\bm p'}$ is in general a function of momenta $\bm q,\bm p$ and $\bm p'$, which to the leading order can be treated as momentum-independent $\beta_{\bm q\bm p\bm p'}=\beta N_0$ if it does not change sign in the region where we are interested. One can work out $\beta$ numerically
\begin{equation}
\beta =-\frac{T}{2N_0}\sum_{n\in\mathbb{Z}}\int d^2\bm k{\rm Tr}[G_e(\bm k,i\omega_n)(i\sigma_y)G_h(\bm k,i\omega_n)(i\sigma_y)^{\dagger}]^2
\end{equation}
by Mastubara Green's functions $G_e=[i\omega_n -\mathcal{H}_{\bm k}]^{-1}$ and $G_h=[i\omega_n +\mathcal{H}^{*}_{-\bm k}]^{-1}$ with $\omega_n=(2n+1)\pi T$.

In the absence of SOC,  quartic coefficient $\beta$ in Eq.  (6) of the main text accidentally vanishes at the upturning point, making it a tricritical point where BCS ($q=0$), FF (single-$\bm q$) and LO ($\pm\bm q$) phases compete.
With SOC considered in this work, $\beta$ is finite as long as SOC is nonzero, and the supercurrent can be calculated by Eq. (11) of the main text. 
In the following we assume SOC is finite such that $\beta\neq 0$ and expansion Eq. (\ref{eq_f1bs}) also applies.

\begin{figure*}[ht]
\begin{center}
\leavevmode\includegraphics[width=1\hsize]{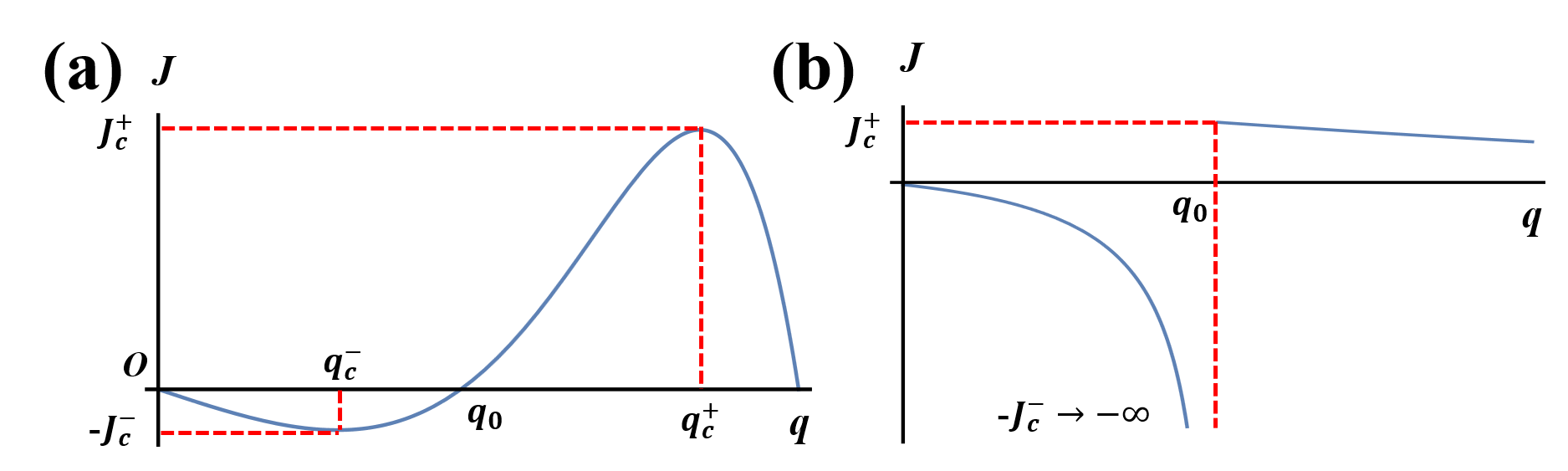}
\end{center}
\caption{(a) Near the FF transition, supercurrent $J$ as a function of Cooper pair momentum $q$ and $J(-q)=-J(q)$. Here $q_{0}$ is the momentum of FF phase, and $q_{c}^{\pm}$ are momenta for critical currents $\pm J_c^{\pm}$ respectively in the positive branch $q>0$. (b) At $T=0$,  supercurrent $J$ as a function of Cooper pair momentum $q$ and $J(-q)=-J(q)$. Here $q_{0}$ is the momentum of FF phase,  and $q_{c}^{\pm}=q_0^{\pm}$ are momenta for critical currents $\pm J_c^{\pm}$ respectively in the positive branch $q>0$. }
\label{fig_S}
\end{figure*}

As shown in Fig. \ref{fig_S}, along a given direction, there are five zeros of supercurrent, one at metastable BCS phase $q=0$, one at ground state FF phase $ q_{0}$, one at opposite FF phase $-q_{0}$ and the other two at excited states where superconductivity vanishes $\alpha_{\bm q}=0$. We hence expect four extremal points $q_c^{\pm},-q_c^{\pm}$ of the supercurrent, and the resulting maximum and minimum are critical currents $\pm J_{c}^{\pm}$ respectively.

Near the metastable BCS phase, the function $\alpha_{\bm q}=\alpha_{-\bm q}$ is fully symmetric, and there seems no diode effect. However, around the true ground state FF phase $\bm q=\bm q_{0}$ we have the expansion up to the nonzero third order
\begin{equation}\label{eq_f2b}
\alpha_{\bm q+\bm q_{0}}={c}_{0}-\frac{c_1^2}{4c_2} +4 c_2\left\{ (\bm q\cdot{\bm q}_{0})^2+ (\bm q\cdot{\bm q}_{0})q^2\right\}.
\end{equation}
Since $|\bm q_{0}|\propto{\rm Re}\sqrt{B^2-B_*^2}$, near the upturning point the third order term is more important than the second order one. Consequently, $\alpha_{\bm q}$ is highly asymmetric near $\bm q_{0}$, and $J_{c}^{\pm}$ can be very different. 

Moreover, as we approach the upturning point, $J_{c}^{-}\to 0$ if superconducting phase stays in the $q>0$ branch. In this case, the superconductor near FF transition is a perfect diode: Supercurrent cannot pass antiparallel to the Cooper pair momentum.
To be precise, the diode coefficient for supercurrent along direction $\pm\hat{\bm n}$ is
\begin{eqnarray}\label{eq_f3b}
\delta_{\hat{\bm n}}=(\hat{\bm n}\cdot\hat{\bm q}_{0})F(x),
\end{eqnarray}
where $x=c({B^2-B_{*}^2})/\sqrt{c_0c_2}$.  
The special function is
\begin{eqnarray}
F(a)=\frac{J_1(a)-J_2(a)}{J_1(a)+J_2(a)},
\end{eqnarray}
where
\begin{eqnarray}
J_1(a)=(ax-2x^3)(x^4-ax^2-1),\\
J_2(a)=-(ay-2y^3)(y^4-ay^2-1)
\end{eqnarray}
and ($A=\sqrt{28+11a^2}$, $\phi=\arccos\frac{20a^3+112a}{A^3}$)
\begin{eqnarray}
x=\sqrt{\frac{1}{7}A\cos\left(\frac{\phi}{3}\right) +\frac{5 a}{14}},\\
y=\sqrt{\frac{1}{7}A\cos\left(\frac{\phi-2\pi}{3}\right) +\frac{5 a}{14}}.
\end{eqnarray}
Notice that $F(x)=0$ for $x<0$, and $F(x)=0.90x^{\frac{3}{2}}-1$ for $0<x\ll 1$.
As a result, when $B=B_*$ and $T<T_*$, we have $x=0$ and the FF superconductor is a perfect diode $\delta =-1$.

It is also possible that near the upturning point, when injected supercurrent switches its direction, superconducting phase changes from the $q>0$ branch to the $q<0$ branch. In that case there is no supercurrent diode effect. Similar discussions can also be found in Ref. \cite{Samokhin}.

\section{Supercurrent diode effect near zero temperature}
Near the transition between FF superconductor and normal phase at low temperature $T\rightarrow 0$, we have  
\begin{equation}
\phi \left(\frac{M}{2\pi T}\right)\sim -\log\frac{T}{T_c}+\log\left|\frac{M}{\Delta_0}\right|
\end{equation}
where $\Delta_0= 4\pi e^{-\psi(1/2)} T_c$ is the pairing gap at zero temperature and zero field. Thus in the absence of SOC, 
\begin{equation}
\alpha_{\bm q}=\frac{1}{2}\sum_{\lambda=\pm}\int_{0}^{2\pi}\frac{d\varphi}{2\pi} \left\{\log\left|\frac{B+\lambda v_{\rm F}q\cos\varphi}{\Delta_0}\right|\right\}
\end{equation}
which can be worked out as $\alpha_{\bm q}=\log(v_{\rm F}q/\Delta_0)$ when $v_{\rm F}q>B$, and in general can be written as the following piecewise function
\begin{equation}\label{eq_f1c}
\alpha_{\bm q}={\rm Re} \left[\log\left(\frac{B+\sqrt{B^2-v_{\rm F}^2q^2}}{\Delta_0}\right) \right].
\end{equation}
Minimizing $\alpha_{\bm q}$ over $\bm q$ yields a large Cooper pair momentum $q_0=B/v_{\rm F}$ approaching $1/\xi_0$ as $B\rightarrow B_c$.  In this case, due to the non-analytic dependence of $\bm q$, $\alpha_{\bm q}$ is highly skewed with respect to ${\bm q}_0$: it rises steeply as $q$ decreases from $q_0$. This leads to the maximum possible  diode effect with $\delta=(J_c^+-J_c^-)/(J_c^++J_c^-) \sim -1$ near $B_c$, taking opposite sign as the one near tricritical point.

\end{document}